# Observation of quasi-steady dark excitons and gap phase in a doped semiconductor

Shangkun Mo[1#], Yunfei Bai[2,3#], Chunlong Wu[1], Xingxia Cui[4], Guangqiang Mei[4], Qiang Wan[1], Renzhe Li[1], Cao Peng[1], Keming Zhao[1], Dingkun Qin[1], Shuming Yu[1], Hao Zhong[1], Xingzhe Wang[1], Enting Li[1], Yiwei Li[1], Limin Cao[4], Min Feng[4], Sheng Meng[2,3,5*], and Nan Xu[1,6*]

[1] *Institute for Advanced Studies, Wuhan University, Wuhan 430072, China*
[2] *Beijing National Laboratory for Condensed Matter Physics and Institute of Physics, Chinese Academy of Sciences, Beijing 100190, China*
[3] *School of Physical Sciences, University of Chinese Academy of Sciences, Beijing 100190, China*
[4] *School of Physics and Technology, Wuhan University, Wuhan 430072, China*
[5] *Songshan Lake Materials Laboratory, Dongguan, Guangdong 523808, China*
[6] *Wuhan Institute of Quantum Technology, Wuhan 430206, China*
[*] E-mails: nxu@whu.edu.cn; smeng@iphy.ac.cn

**Excitons drive optical phenomena and correlated phases. Unlike bright excitons, dark excitons evade conventional optical detection, and their electronic modulation effects in quasi-equilibrium remain elusive. Here, using angle-resolved photoemission spectroscopy, we report creating, detecting, and controlling dark excitons in the quasi-equilibrium distribution in a doped semiconductor $SnSe_2$. We further observe an anisotropic gap for the conduction band, which is directly related to dark excitons. Our results broaden the scope of dark excitons, extending their studies from the picosecond timescale in the ultrafast photoemission process to conditions occurring under quasi-equilibrium. We reveal a pioneering type of light-matter interaction in the engineering of electronic structures and provide a way to realize the excitonic gap phase in semiconductors with large band gaps.**

The concept of quasi-particles allows us to reveal complex many-body interactions in condensed matter and understand various unconventional phenomena. Exciton, the bound state of an electron-hole pair, can be excited by light in semiconductors and considerably affect the fundamental properties of semiconductor devices [1,2]. Furthermore, macroscopic quantum phenomena such as excitonic insulator (EI) [3-6], charge order wave (CDW) [7-9], and superfluidity [10] can be induced by low to intermediate exciton densities at low temperatures with the coexistence of electrons and quasiparticle holes.

The bright exciton is optically active and can thus be directly detected by optical spectroscopy techniques [11-13]. Indirect exciton semiconductors exhibit momentum-forbidden optical transitions, rendering dark excitons undetectable by conventional optical methods without phonon assistance [14]. Dark exciton not only plays an essential role in understanding dynamics in semiconductors [15], but also has a longer lifetime for promoting high-temperature macroscopic quantum phenomena and excitonic transport [16]. Recent breakthroughs in time- and angle-resolved photoemission spectroscopy (TR-ARPES) with extreme ultraviolet photons allow the direct visualization of dark excitons and study of their dynamics [14,17-21]. However, experimental detection of quasi-steady dark excitons and their band structure modulation remains experimentally challenging, despite their critical role in semiconductor device functionalities. Besides, excitonic insulating phase has so far only been discussed in systems near the semimetal-semiconductor transitions, in which the value of band gap $E_g$ has to be smaller than the exciton binding energy $E_b$ [3-6].

Here, we experimentally demonstrate that angle-resolved photoemission spectroscopy (ARPES) is a direct probe of dark excitons in the quasi-equilibrium distribution. ARPES provides photo-generated holes for forming dark exciton in electron-doped semiconductor SnSe$_2$ and simultaneously probes the electron-hole pairs as evidenced by valence band replicas right below conduction bands in the gap region. We extract binding energy $E_b \sim 0.5 \text{ eV}$ and Bohr radius $R_B \sim 0.4 - 0.6 \text{ nm}$. Meanwhile, we observe an energy gap gradually opening near $E_\text{F}$ for the conduction band as exciton is established. By increasing temperature, dark excitons and gaps gradually disappear in a simultaneous way. We demonstrate light-induced quasi-steady dark excitons and excitonic gap phases in doped SnSe$_2$, establishing a novel platform to probe and manipulate correlated exciton physics.

High-quality single crystals of 1T-SnSe$_2$ were synthesized using the chemical vapor deposition method [22]. The potassium (K) atoms were in situ deposited on clean surfaces at T = 6 K. ARPES measurements were conducted at a home-designed facility with $hv = 21.2$ eV. Data were acquired under a vacuum better than $5 \times 10^{-11}$ mbar with energy/angular of 22 meV and 0.1°. $E_\text{F}$ was determined by measuring a sputtered copper reference in electrical contact with the samples. No surface charging effect is observed in the entire doping range.

Figure 1 illustrates the mechanism of probing quasi-steady exciton with ARPES. As seen from Fig. 1a, photo-generated holes are excited by the incident light in the ARPES experiment (inset (i)), and electron carriers are induced by either surface K deposition (inset (ii)) or Se-vacancy (inset (iii)). The doped electrons and photo-generated holes form excitons (inset (iv)), which have long lifetimes due to the indirect gap nature. The density of photo-generated hole in quasi-equilibrium, as estimated in (Supplemental Material (SM) Sec. S2), is way lower than that in ultrafast experiments [17,23]. The incident light can further ionize the electron of an exciton, generating a dressed electron in the photoemission process (v). The dressed photoelectron manifests as the replica of valence band dispersion located below the conduction band minimum, displaced by the energy of $E_b$, as theoretically discussed in previous work [24].

Figure 1b displays the experimentally determined band structure of pristine SnSe$_2$, where the three valence bands α, β and γ are observed, consistent with previous results [25]. The first-principles calculations are appended, that agree well with the ARPES results. We in situ deposit K atoms on the SnSe$_2$ surface, and the conduction band shifts below $E_\text{F}$ around the $\overline{\text{M}}$ point (Fig. 1c). Furthermore, we observe an additional spectral weight within the band gap that is absent in first-principles calculations. The additional spectral weight is relatively weak compared to both valence and conduction bands, therefore intensity in corresponding region is amplified by a factor of 3 in Fig. 1c. The additional spectra weight shows a similar dispersion to the valence β, as the direct signature of quasi-steady dark excitons. In Fig. 1d, we plot the simulation of ARPES spectra of exciton with a narrow distribution of center-of-mass momentum Q = 0, which well reproduce the experimental observations in Fig. 1c. The exciton binding energy is determined as $E_b = 480$ meV by the integrated energy distribution curve (EDC) at the $\overline{M}$ point in Fig. 1e. Both the $E_b$ and energy broadening of exciton signature in SnSe$_2$ are comparable to those in WSe$_2$/WS$_2$ monolayers [26,27]. Our theoretical calculations of SnSe$_2$ monolayer excitonic properties reveal qualitative agreement between the Q = 0 exciton with a narrow distribution and experimental observations, attributable to the 2D character of both photo-generated holes and surface doping-induced electron carriers (see SM Sec. S4). We also note that the upper valence band α shows a strong three-dimensional (3D) character (see SM Sec. S5), instead of 2D behavior of the β band. The experimental signature in Fig. 1c corresponds to excitons with large binding energy, which strongly suggests a 2D character with binding energy enhancement [13,28]. Consequently, the replica band observed at $\overline{M}$ point in Fig. 1c exhibits a dispersion that closely mimics that of the 2D β band, instead of the 3D α band.

We can examine exciton-induced replicas from the perspective of momentum distribution and give an estimate of the size of exciton in real space (Fig. 1f). In Fig. 1g, we plot ARPES map integrated with a binding energy window $0.3 \text{ eV} < E_B < 1 \text{ eV}$, indicating a strong anisotropic momentum distribution of exciton-induced replica. Our calculations of the exciton momentum distribution in Fig. 1h are fully support experimental results. It can also be clearly seen that the replica is more confined along the $\overline{K} - \overline{M} - \overline{K}$ direction in Fig. 1i than that along the $\overline{\Gamma} - \overline{M} - \overline{\Gamma}$ direction in Fig. 1c. We next fit the integrated momentum distribution curve (MDC) with an excitonic wave function squared $|\emptyset(k)|^2 \propto 1/[1 + (kR_B)^2/4)]^4$, where $R_B$ is the exciton Bohr radius, yields $R_B = 6.4$ Å along the $\overline{K} - \overline{M} - \overline{K}$ direction (Fig. 1k). By fitting a series of integrated MDCs, we obtain an elliptic exciton distribution and estimate an anisotropic ratio of 1.64 for the exciton size in real space, which is in a good agreement with our theoretical calculations. (see SM Sec. S9).

As the exciton-induced replica band appears, we observe a global energy gap near $E_\text{F}$ for the conduction band near the $\overline{\text{M}}$ point (Fig. 2a). As seen from the high-resolution cut along the $\overline{\text{M}} - \overline{\text{K}}$ direction (cut 1 in Fig. 2b) in Fig. 2c-I, spectra weight of conduction band suddenly drops near $E_\text{F}$. Figure 2d-I displays symmetrized EDCs at Fermi momentums ($k_\text{F}$) [29,30] of the conduction band (see SM Sec. S13). From the double peaks feature, we can determine a gap opening with $\Delta = 90$ meV along cut1 direction. In order to obtain the overall gap structure, we measured a series of high-resolution cuts off the high symmetry line (indicated in Fig. 2b), with band structures and symmetrized EDCs plotted in Fig. 2c and 2d. As moving away from the $\overline{\text{M}}$ point, the gap value becomes smaller, with the minimum value of gap of $\Delta_{min} = 65$ meV located along the $\overline{\Gamma} - \overline{\text{M}}$ direction. The gap structure in k-space is summarized in Fig. 2b. The anisotropic gap structure excludes the impurity scattering scenario for the gap opening.

To gain further insights into the nature of the dark excitons, we tune electron carrier density by controlling the surface K deposition dose and examine the evolution of exciton band and energy gap near $E_\text{F}$. With longer K evaporating time, more electron carrier density is induced as evidenced by increases in conduction band filling (Fig. 3a). In the meantime, the excitonic features in photoemission spectra become increasingly prominent (e.g., the spectral weight of excitons in Fig. 3a is weaker compared to Fig. 1c), which can be quantitatively analyzed by the spectra area of the hump feature in the integrated EDCs of different doping levels (Fig. 3c). The intensity increase of replica band with more electron carrier density is consistent with exciton scenario described in Figs. 1a-b, because more electrons doped on surface form more electron-hole pairs, which induce more exciton density and corresponding hump spectra weight. The gap value near $E_\text{F}$ also increases with more electron carrier density (Fig. 3b), where the gap value $\Delta$ variation extracted by symmetrized EDC at $k_\text{F}$ in Fig. 3d. The doping level dependences of excitonic effect and gap size $\Delta$ are summarized in Fig. 3e, which follow the same trend. The electron carrier density is estimated by conduction FS volume (see SM). We note that, with electron doping, both the conduction band and exciton band shift downward from $E_\text{F}$, and consequently the exciton binding energy $E_b$ remains unchanged. This behavior excludes the plasmon scenario for the replica band because the plasma energy becomes larger with higher carrier density [31,32] (see SM Sec. S7 for more doping details).

Because of K-atoms desorption from the surface at rising temperature, we turn to SnSe$_{1.9}$ samples, in which Se vacancies provide electron carriers instead of surface K doping, to examine the temperature-dependent behavior. We also observe excitonic phenomena in SnSe$_{1.9}$ at T = 6 K, with the same mechanism as that in K-doped SnSe$_2$. As seen from the band structure of SnSe$_{1.9}$ in Fig. 4a, the conduction band is observed below $E_F$ due to the doping effect induced by Se vacancies, and a similar valence band replica appears below the conduction band indicating the dark exciton. We noted that the bulk SnSe$_{1.9}$ sample is doped by Se vacancies, which enhances the screening effect of Coulomb interactions compared to the case of 2D surface-doped SnSe$_2$. It consequently induces a slightly smaller exciton binding energy $E_b = 310$ meV in SnSe$_{1.9}$ than that in the surface-doped SnSe$_2$. We also observe a gap opening near $E_F$ in SnSe$_{1.9}$, with gap size $\Delta = 43$ meV at T = 6 K (Fig. 4b). As increasing temperature, the hump corresponding to exciton-induced replica gradually suppressed (Fig. 4c), in the meantime, the gap near $E_F$ gradually closes (Fig. 4d). We extract the temperature dependences of exciton hump area and gap value in Fig. 4e, which follow the same trend. We can fit the gap value $\Delta(T)$ with the BCS mean-field equation, yielding a gap closing temperature $T_c = 80$ K. Our results demonstrate a striking correlation strength $\frac{2\Delta}{k_B T_c} = 12.6$, consistent with the strong correlation nature of excitonic physics. The exciton disappears at a slightly higher temperature T ~ 100 K than the gap closing temperature $T_c$.

In summary, our conventional ARPES measurements directly access quasi-steady dark excitons in doped SnSe$_2$, with observed valence band replicas within the band gap agreeing with theoretical excitonic photoemission signatures [24]. Our calculations of exciton properties are fully consistent with that determined by ARPES results, further supporting the excitonic scenario. We can clearly rule out other electron-bosonic mode coupling mechanisms for the experimental observations of peak-dip-hump feature, such as phonon/polaron [33-35], magnon [36], and plasmon [31,32,37]. Generally, in the cases of electron-bosonic mode coupling, the hump features of EDC correspond to conduction band replica [33-35], instead of a valence band replica observed here [24]. The large energy-momentum separation between valence band and its replica definitively excludes bosonic modes: (i) phonon coupling is incompatible with the observed energy scale [38], (ii) magnons are forbidden by the nonmagnetic nature of SnSe$_2$, and (iii) doping dependence rules out plasmons.

We observe an anisotropic conduction band gap opening near $E_\text{F}$, exhibiting doping- and temperature-dependent evolution correlated with quasi-steady dark exciton intensity, suggesting a direct excitonic origin. The gap phase observed here shows distinction from that in exciton insulator phases in semimetals and semiconductors, due to a different exciton formation mechanism (Figs. 5a-c). For exciton insulator phases in semimetals or narrow gap semiconductors with band gap $E_g < E_b$, electrons and holes within the energy window of $E_b$ are spontaneously paired at low temperatures, resulting in reconstituted bands with an enhanced gap value fixed at $\Delta = E_b$ (Figs. 5a-b). In this case, the exciton binding energy $E_b$ (or gap value $\Delta$) reaches its maximum value when the electron and hole densities are equal, as the screening effect of free carriers is weak [39]. However, in this work, the excitons are formed by electrons near $E_\text{F}$ and photo-generated holes of valence bands (Fig. 5c). The energy of photo-generated holes is well below $E_\text{F}$ and falls out of the energy window of $E_b$. The observed gap phase likely stems from exciton-driven band renormalization, forming a many-body gap akin to Mott physics. This gap emerges via photo-generated hole-mediated electron-electron interactions, paralleling mechanisms in superconductivity. Our findings expand the horizons of dark excitons in light-matter interaction in the modulation of electronic structures while introducing an innovative approach to engineer excitonic gap phase in large band gap semiconductors. This methodology offers new possibilities for manipulating electronic states in surface electrical transport measurements.

**Acknowledgement**

We thank the support with the ARPES experiments by BL03U beamline of the Shanghai Synchrotron Radiation Facility.

**Data availability**

All data are processed by Igor Pro 8.0 software. All data generated during the current study are available from the corresponding author upon request.

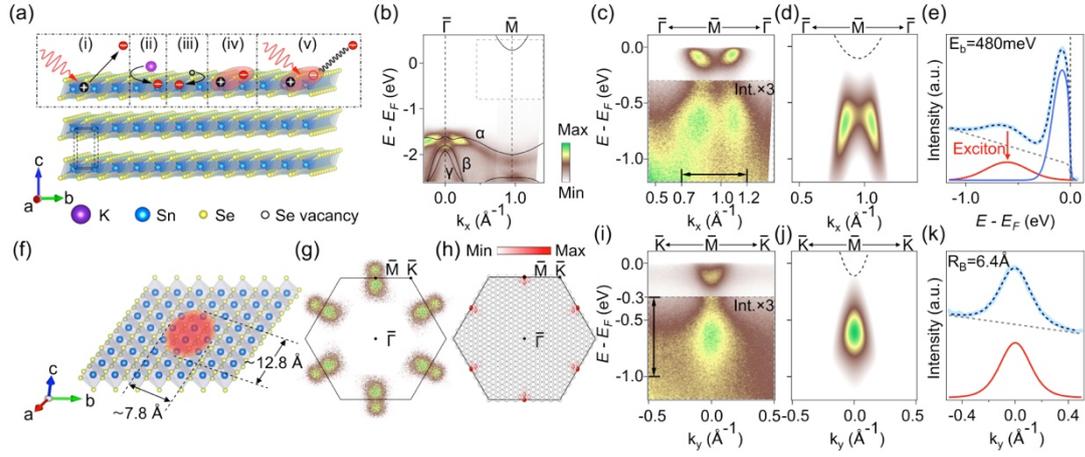

**Fig. 1.** (a) Schematic of the formation of photo-induced dark exciton in ARPES experiments. (b) Band structures of pristine SnSe$_2$ along the $\bar{\Gamma} - \bar{M}$ direction, with First-principle calculations appended. (c) ARPES intensity plot of surface doped SnSe$_2$ along the $\bar{\Gamma} - \bar{M}$ direction near the $\bar{M}$ point. (d) Corresponding ARPES spectra simulation. (e) Integrated EDC within momentum window indicated by the horizontal arrow in (c) and fitting analysis. (f) Schematic of exciton in real space. (g) The $k_x - k_y$ ARPES map near the $\bar{M}$ point integrated within an energy window indicated by the vertical arrow in (i). (h) Calculated momentum distribution of the electrons forming excitons. (i)-(j) Same as (c)-(d) but along the $\bar{M} - \bar{K}$ direction. (k) MDC integrated within the same energy window as (g). The electron density is estimated to be 0.086 per unit cell for (c)-(k).

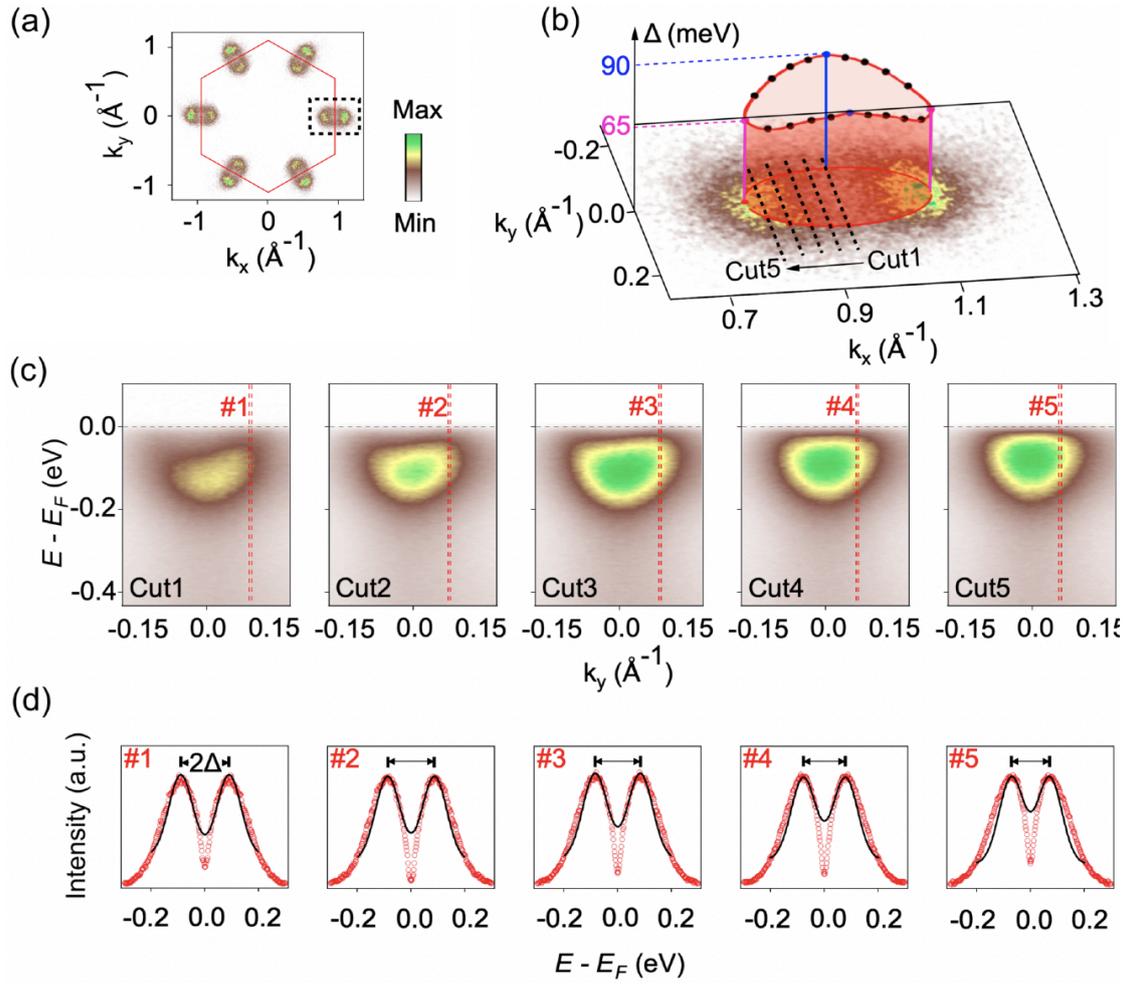

**Fig. 2.** (a) The $k_x - k_y$ ARPES map near $E_F$ integrated within an energy window of 0.1 eV. (b) Momentum resolved gap structure. (c) Band structures near $E_F$ along cuts 1-5. (d) Corresponding symmetrized EDCs #1-#5 at $k_F$, with fitting analysis. The electron density is estimated to be 0.086 per unit cell.

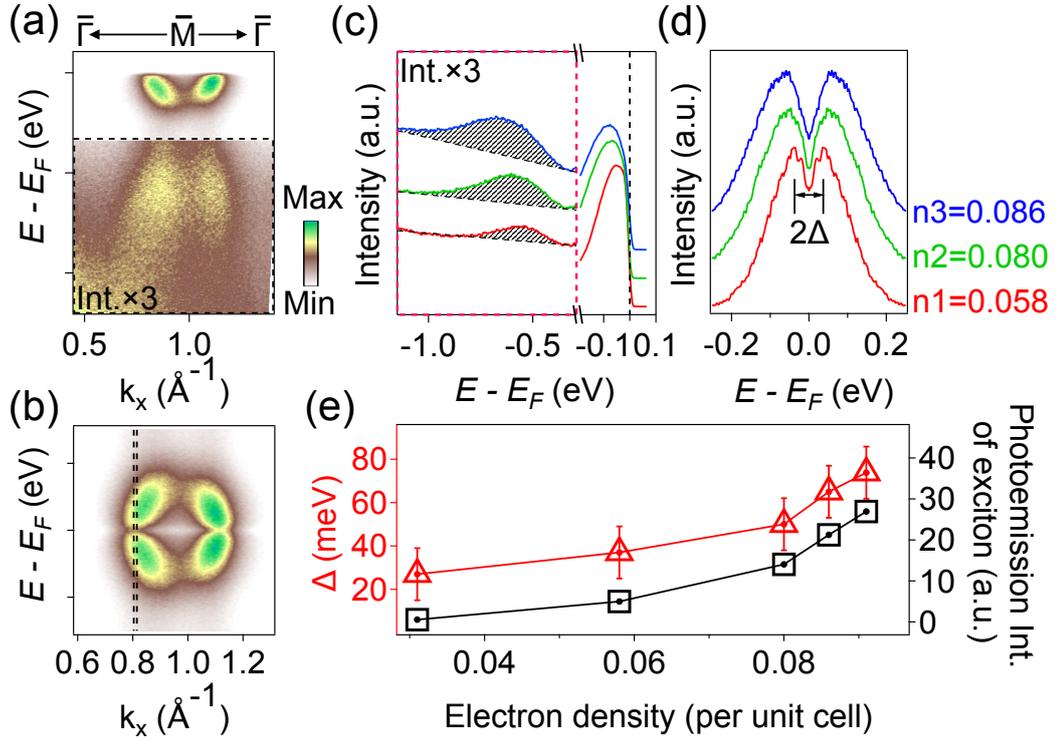

**Fig.3.** (a) The ARPES intensity plot along the $\bar{\Gamma} - \bar{M}$ direction with an electron density 0.080 per unit cell. (b) Corresponding symmetrized ARPES intensity plot with respect to $E_F$. (c) Integrated doping-dependent EDCs near the $\bar{M}$. (d) Doping-dependent symmetrized EDCs taken at $k_F$. The electron densities in (c)-(d) are estimated to be 0.058, 0.080, 0.086 per unit cell, respectively. (e) The energy gap $\Delta$ and the intensity of exciton, as a function of electron density, respectively.

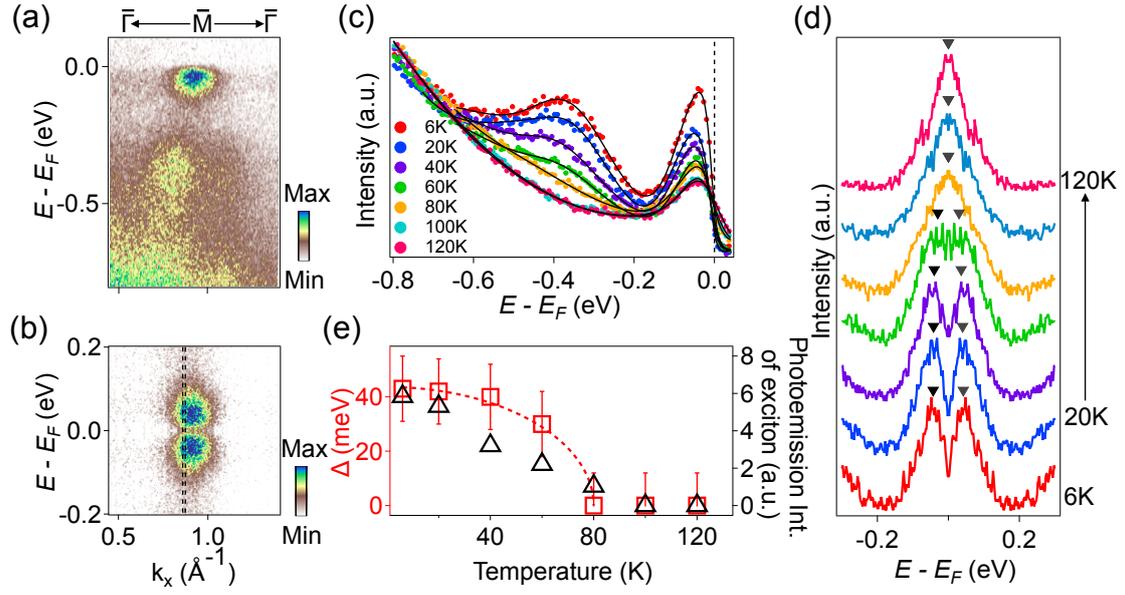

**Fig. 4.** (a) ARPES intensity plots along the $\bar{\Gamma} - \bar{M}$ direction at T = 6 K. (b) Corresponding symmetrized ARPES intensity plot with respect to $E_F$. (c) Temperature dependent integrated EDCs measured at the $\bar{M}$ point. (d) Temperature dependence of symmetrized EDCs taken at $k_F$. (e) The evolutions of the gap and the intensity of exciton with temperature.

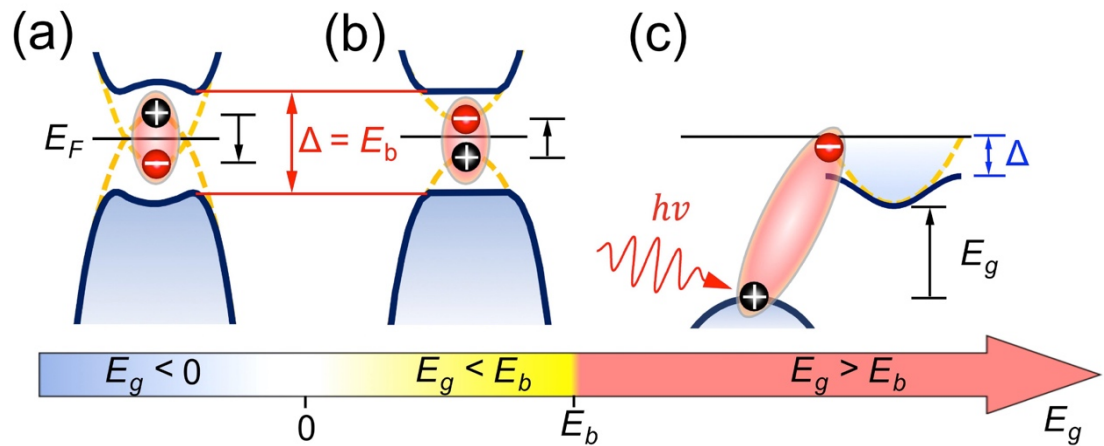

**Fig. 5.** (a) Exciton insulating phase in a semimetal with negative $E_g$ (black arrow). The normal states are represented by dashed parabolas. (b) Same as (a), but for narrow gap semiconductor with $E_g < E_b$. In both cases, the exciton-induced gap value marked by the red arrow is equal to the exciton binding energy $E_b$. (c) Light-induced excitonic gap phase in a semiconductor with $E_g > E_b$. The gap value near $E_F$ is denoted by the blue arrow.

**References**

[1] Ross, J. S. et al. Electrically tunable excitonic light-emitting diodes based on monolayer $WSe_2$ p–n junctions. Nat. Nanotechnol. 9, 268–272 (2014).

[2] Mak, K. F. & Shan, J. Photonics and optoelectronics of 2D semiconductor transition metal dichalcogenides. Nat. Photon. 10, 216–226 (2016).

[3] Mott, N. F. The transition to the metallic state. Philos. Mag. A J. Theor. Exp. Appl. Phys. 6, 287–309 (1961).

[4] Jérome, D., Rice, T. M. & Kohn. W. Excitonic insulator. Phys. Rev. 158, 462–475 (1967).

[5] Du, L. et al. Evidence for a topological excitonic insulator in InAs/GaSb bilayers. Nat. Commun. 8, 1971 (2017).

[6] Jia, Y., Wang, P., Chiu, CL. et al. Evidence for a monolayer excitonic insulator. Nat. Phys.18, 87–93 (2022).

[7] Cercellier, H. et al. Evidence for an excitonic insulator phase in $1T-TiSe_2$. Phys. Rev. Lett. 99, 146403 (2007).

[8] Gao, Q., Chan, Yh., Jiao, P. et al. Observation of possible excitonic charge density waves and metal–insulator transitions in atomically thin semimetals. Nat. Phys. 20, 597-602 (2024).

[9] Song, Y., Jia, C., Xiong, H. et al. Signatures of the exciton gas phase and its condensation in monolayer $1T-ZrTe_2$. Nat. Commun. 14, 1116 (2023).

[10] Fogler, M., Butov, L. & Novoselov, K. High-temperature superfluidity with indirect excitons in van der Waals heterostructures. Nat. Commun. 5, 4555 (2014).

[11] Wang, G., Chernikov, A., Glazov, M. M. et al. Colloquium: Excitons in atomically thin transition metal dichalcogenides. Rev. Mod. Phys. 90, 021001 (2018).

[12] Fukutani, K., Stania, R., Il Kwon, C. et al. Detecting photoelectrons from spontaneously formed excitons. Nat. Phys. 17, 1024–1030 (2021).

[13] Ma, J., Nie, S., Gui, X. et al. Multiple mobile excitons manifested as sidebands in quasi-one-dimensional metallic $TaSe_3$. Nat. Mater. 21, 423–429 (2022).

[14] Wallauer, R. et al. Momentum-Resolved Observation of Exciton Formation Dynamics in Monolayer $WS_2$. Nano Lett. 21, 5867–5873 (2021).

[15] Chand, S.B., Woods, J.M., Quan, J. et al. Interaction-driven transport of dark excitons in 2D semiconductors with phonon-mediated optical readout. Nat. Commun. 14, 3712 (2023).

[16] Unuchek, D. et al. Room-temperature electrical control of exciton flux in a van der Waals heterostructure. Nature 560, 340–344 (2018).

[17] Madéo, J. et al. Directly visualizing the momentum-forbidden dark excitons and their dynamics in atomically thin semiconductors. Science 370, 1199 (2020).

[18] Schmitt, D., Bange, J.P., Bennecke, W. et al. Formation of moiré interlayer excitons in space and time. Nature 608, 499–503 (2022).

[19] Karni, O., Barré, E., Pareek, V. et al. Structure of the moiré exciton captured by imaging its electron and hole. Nature 603, 247–252 (2022).

[20] Mori, R., Ciocys, S., Takasan, K. et al. Spin-polarized spatially indirect excitons in a topological insulator. Nature 614, 249–255 (2023).

[21] Dong, S. et al. Direct measurement of key exciton properties: Energy, dynamics, and spatial distribution of the wave function. Nat Sci. 1 e10010 (2021).

[22] Jiao, C. X, Huang. K. et al., Resonant four-photon photoemission from $SnSe_2$ (001). Frontiers of Physics 19, 33207 (2024).

[23] Park, H., Zhu, J., Wang, X. et al. Dipole ladders with large Hubbard interaction in a moiré exciton lattice. *Nat. Phys.* 19, 1286–1292 (2023).

[24] Rustagi, A. Kemper, A. F. Photoemission signature of excitons. Phys. Rev. B 97, 235310 (2018).

[25] Pham, A. T. et al. High-quality $SnSe_2$ single crystals: Electronic and thermoelectric properties. ACS Appl. Energy Mater. 3, 11, 10787–10792 (2020).

[26] Man, M. K. L. et al. Experimental measurement of the intrinsic excitonic wave function. Sci. Adv. 7, eabg0192 (2021).

[27] Chernikov, A. et al. Exciton binding energy and nonhydrogenic rydberg series in monolayer $WS_2$. Phys. Rev. Lett. 113, 076802 (2014).

[28] Gonzalez, J. M. Oleynik, I. I. Layer-dependent properties of $SnS_2$ and $SnSe_2$ two-dimensional materials. Phys. Rev. B 94, 125443 (2016).

[29] Norman, M., Ding, H., Randeria, M. et al. Destruction of the Fermi surface in underdoped high-Tcsuperconductors. Nature 392, 157–160 (1998).

[30] Norman, M. R. et al. Phenomenology of the low-energy spectral function in high- Tc superconductors. Phys. Rev. B 57, R11093 (1998).

[31] Riley, J. M., Caruso, F., Verdi, C. et al. Crossover from lattice to plasmonic polarons of a spin-polarised electron gas in ferromagnetic EuO. Nat. Commun. 9, 2305 (2018).

[32] Ulstrup, S., in 't Veld, Y., Miwa, J.A. et al. Observation of interlayer plasmon polaron in graphene/$WS_2$ heterostructures. Nat. Commun. 15, 3845 (2024).

[33] Lee, J. J. et al. Interfacial mode coupling as the origin of the enhancement of $T_C$ in FeSe films on $SrTiO_3$. Nature 515, 245–248 (2014).

[34] Wang, Z. et al. Tailoring the nature and strength of electron-phonon interactions in the $SrTiO_3$(001) 2D electron liquid. Nat. Mater. 15, 835–839 (2016).

[35] Moser, S. et al. Tunable polaronic conduction in anatase $TiO_2$. Phys. Rev. Lett. 110, 196403 (2013).

[36] Dahm, T. et al. Strength of the spin-fluctuation-mediated pairing interaction in a high-temperature superconductor. Nat. Phys. 5, 217–221 (2009).

[37] Bostwick, A. et al. Observation of plasmarons in quasi-freestanding doped graphene. Science 328, 999–1002 (2010).

[38] Wang, H. F. et al. Anisotropic phonon transport and lattice thermal conductivities in tin dichalcogenides $SnS_2$ and $SnSe_2$. RSC Adv., 7, 8098 (2017).

[39] Ma, L., Nguyen, P.X., Wang, Z. et al. Strongly correlated excitonic insulator in atomic double layers. Nature 598, 585–589 (2021).